\documentclass{article}
\usepackage{amssymb}
\usepackage{multirow}
\usepackage{graphicx}
\usepackage{spconf,amsmath,graphicx}
\usepackage{mathrsfs}
\usepackage{bbding}
\usepackage{cite}
\usepackage[urlcolor=blue]{hyperref}

\title{Multi-view MidiVAE: Fusing Track- and Bar-view Representations for Long Multi-track Symbolic Music Generation}
\address{Author Affiliation(s)}
\name{
\begin{tabular}{c}
Zhiwei Lin$^{1,2,*}$, Jun Chen$^{1,3,*}$\thanks{*These authors contributed equally to this work as first authors.}, Boshi Tang$^1$, Binzhu Sha$^1$, Jing Yang$^2$, Yaolong Ju$^2$, \\ \textit{Fan Fan$^2$, Shiyin Kang$^3$, Zhiyong Wu$^{1,4,\dagger}$\thanks{$\dagger$ Corresponding author.}, Helen Meng$^{4}$}
\end{tabular}
\vspace{-0.2cm}
}

\address{$^1$
    Shenzhen International Graduate School, Tsinghua University, Shenzhen, China \\
  $^2$ Huawei Technologies Co., Ltd., Shenzhen, China  \quad $^3$Skywork AI PTE. LTD. \\
    $^4$ The Chinese University of Hong Kong, Hong Kong SAR, China\\
    \small{
        \{lzw22, y-chen21\}$@$mails.tsinghua.edu.cn, 
        zywu$@$sz.tsinghua.edu.cn
    }
}
\begin{document}
\ninept 
\maketitle
\begin{abstract}
Variational Autoencoders (VAEs)  constitute a crucial component of neural symbolic music generation, among which some works have yielded outstanding results and attracted considerable attention.
Nevertheless, previous VAEs still encounter issues with overly long feature sequences and generated results lack contextual coherence, thus the challenge of modeling long multi-track symbolic music still remains unaddressed.
To this end, we propose Multi-view MidiVAE, as one of the pioneers in VAE methods that effectively model and generate long multi-track symbolic music.
The Multi-view MidiVAE utilizes the two-dimensional (2-D) representation, OctupleMIDI, to capture relationships among notes while reducing the feature sequences length.
Moreover, we focus on instrumental characteristics and harmony as well as global and local information about the musical composition by employing a hybrid variational encoding-decoding strategy to integrate both Track- and Bar-view MidiVAE features.
Objective and subjective experimental results on the CocoChorales dataset demonstrate that, compared to the baseline, Multi-view MidiVAE exhibits significant improvements in terms of modeling long multi-track symbolic music.

\end{abstract}
\vspace{-0.1cm}

\begin{figure*}[!htbp]
	\centering
    \includegraphics[width=0.8\linewidth]{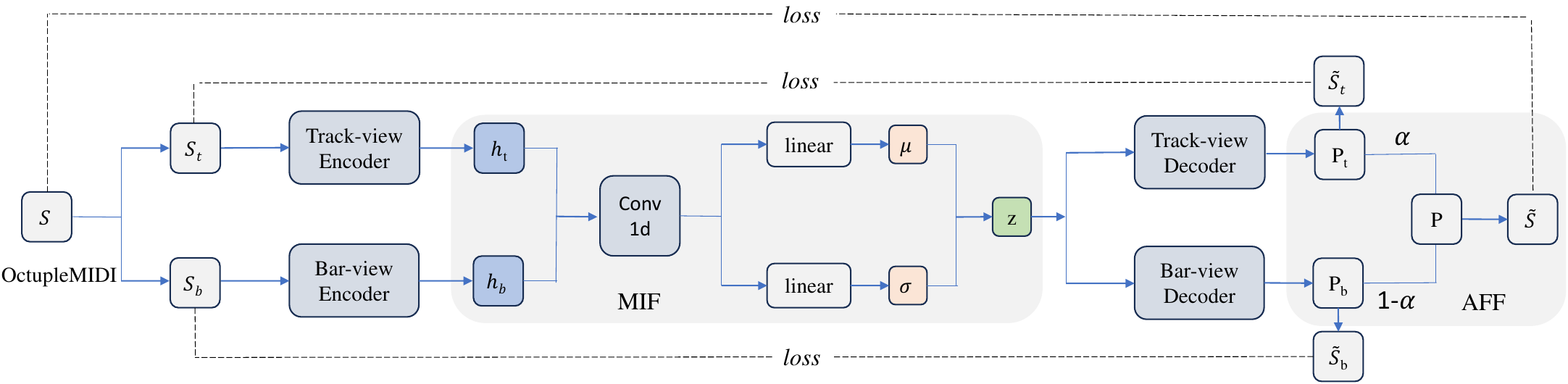}
    \vspace{-0.4cm}
	\caption{The overall diagram of the proposed Multi-view MidiVAE. The model mainly contains Track- and Bar-view encoders, a multi-view information fusion (MIF), Track- and Bar-view decoders as well as an adaptive feature fusion (AFF).}

	\label{fig:totalarch}
	\vspace{-0.5cm}
\end{figure*}

\vspace{-0.3cm}
\section{Introduction}
\vspace{-0.2cm}
\label{sec:intro}
Automatic music generation is a prominent and active research field, 
which has a wide range of applications from human-computer interactive composition to commercial composition software \cite{liu2023wavjourney}.
In recent years, the advancement of deep learning technology further promotes the development of auto music generation.
Generally, automatic music generation can be categorized into symbolic music generation \cite{ren2020popmag,dong2018musegan,dong2023multitrack, huang2018music, donahuelakhnes} and music audio generation \cite{agostinelli2023musiclm,liu2023audioldm,copet2023simple, lam2023efficient, huang2023noise2music, huang2023make}. 
While music audio generation directly models continuous audio signals, symbolic music generation models and synthesizes music as a sequence of symbols, which by nature is more conducive to subsequent music editing and adjustments \cite{ji2023survey}. 
Therefore, symbolic music generation stays an important research topic.

Among symbolic music generation methods, the VAE approaches \cite{roberts2018hierarchical, hadjeres2017glsr, pati2021attribute, wang2020pianotree, wei2022learning} have been widely applied due to their promising performance. 
Additionally, as conditional and multi-modal music generation becomes increasingly popular, VAEs can also serve as an essential component for the extraction and representation of latent features \cite{mittal2021symbolic}.
Consequently, the VAE-based symbolic music generation methods have garnered widespread attention.
MusicVAE \cite{roberts2018hierarchical} employs a Long Short-Term Memory (LSTM)-based VAE architecture to effectively reconstruct and generate symbolic music. 
However, due to the instrument-specific Encoder and Decoder design, MusicVAE is restricted to music generation within designated instruments, which significantly limits its applicability.
In response to this limitation, the Multi-track MusicVAE \cite{simon2018learning} incorporates the MIDI-like \cite{oore2020time} event-based representation that can denote different instruments, and utilizes a universal Encoder and Decoder for modeling. 
This innovation successfully broadens the scope of symbolic music reconstruction and generation from specific instruments to arbitrary instruments, establishing it as one of the most effective VAE-based symbolic music generation methods so far.

Despite the impressive performance, VAE approaches still struggle with the emerging challenge of modeling long multi-track symbolic music.
The MIDI-like \cite{oore2020time} input to the Multi-track MusicVAE is a one-dimensional (1-D) representation neglecting the relationships among notes, which potentially impairs the model's learning for the musical composition.
More critically, converting long multi-track symbolic music into this representation leads to excessively long feature sequences, which substantially increases the burden on model training and inference.
Furthermore, when tackling long multi-track symbolic music, the Multi-track MusicVAE can only process one bar at a time, which results in the final generated music lack of contextual consistency and coherence, severely deteriorating the quality of the symbolic music.

In this paper, to confront the challenges above, we propose Multi-view MidiVAE, a pioneer VAE that efficiently models and generates the long multi-track symbolic music.
In light of the issues associated with the aforementioned MIDI-like features, we introduce a 2-D representation called OctupleMIDI \cite{zeng2021musicbert}. 
This representation takes the relationships among notes into account while reducing the sequence length, subsequently alleviating the complexity of feature extraction and reconstruction of the model.
Moreover, we model from dual views:
1) To focus on instrumental characteristics and harmony as well as the global information of musical composition, we design a transformer-based \cite{vaswani2017attention} Track-view MidiVAE, which adopts an intra- and inter-instrument modeling over the musical composition;
2) Alternatively, the transformer-based Bar-view MidiVAE implements an intra- and inter-bar architecture, concentrating on the finer details of notes within bars.
Eventually, through a hybrid variational encoding-decoding strategy, we integrate the outputs from these two VAE views to form our final Multi-view MidiVAE model.
Objective and subjective experimental results on the CocoChorales dataset \cite{wu2022chamber} validate the efficacy of our improvements, demonstrating that the Multi-view MidiVAE is indeed a potent VAE approach for modeling long multi-track symbolic music.

\vspace{-0.3cm}
\section{Methodology}
\vspace{-0.2cm}
\label{sec:format}
We focus on the task of long multi-track music generation, i.e., generating multi-bar music with various instruments played together.
To this end, we propose Multi-view MidiVAE that effectively models and generates long multi-track symbolic music.
As shown in Fig.\ref{fig:totalarch}, our Multi-view MidiVAE adopts a hybrid variational encoder-decoder architecture, which mainly contains Track- and Bar-view encoders, a multi-view information fusion, Track- and Bar-view decoders as well as an adaptive feature fusion.
More specifically, the input OctupleMIDI sequences $S$ are transformed into track-view sequences $S_t$ and bar-view sequences $S_b$. 
These two sequences are fed into the Track- and Bar-view encoders to extract track-view information contained in feature $h_t$ and bar-view information included in feature $h_b$. 
By means of concatenating $h_t$ and $h_b$, and passing them through a Conv1d layer, we achieve multi-view information fusion and then obtain the hybrid embedding that is utilized to get the mean $\mu$ and variance $\sigma$ of the latent distribution $z$. 
After that, the Track- and Bar-view decoders generate probability matrix $P_t$ and $P_b$. 
Eventually, the adaptive feature fusion fuses $P_t$ and $P_b$ into the final probability matrix $P$ for generating the reconstruction sequences $\widetilde{S}$.
We will elaborate on our improvements in the following subsections.

\begin{figure}[!htbp]
	\centering
    \vspace{-0.2cm}
    \includegraphics[width=0.8\linewidth]{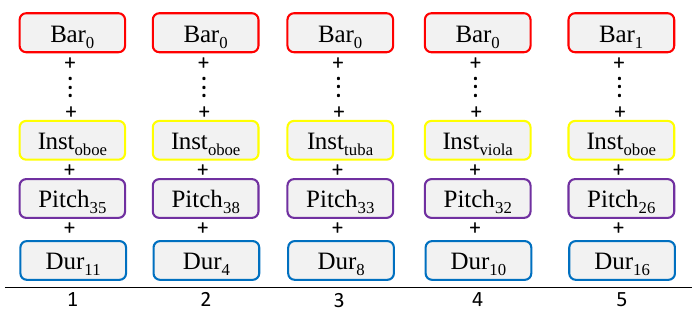}
    \vspace{-0.4cm}
	\caption{The schematic diagram of OctupleMIDI. The ``Dur" and ``Inst" mean duration and instrument respectively.}
	\label{fig:2dpre}
	 \vspace{-0.7cm}
\end{figure}

\subsection{OctupleMIDI Representation}
\vspace{-0.1cm}
Previous 1-D event-based representation  \cite{oore2020time} of Multi-track MusicVAE exhibits certain limitations.
First of all, the representation treats all musical attributes equally, thus limiting the model's ability to understand the differences and connections between attributes.  
Besides, it makes it difficult to capture complex relationships between notes, e.g., accurately representing the simultaneous play of two notes. 
Finally, for processing the long multi-track symbolic music, the 1-D representation increases computational resource and time expenses.

In order to address the aforementioned problems, we adopt the OctupleMIDI representation \cite{zeng2021musicbert}, which consists of sequences of octuple tokens. 
As shown in Fig.\ref{fig:2dpre}, each token consists of eight music attributes, namely pitch, velocity, duration, instrument, position, bar, time signature and tempo. 
The representation employs the note as the fundamental unit and consolidates all musical attributes into a single octuple token.
This enables the model to differentiate music attributes and effectively capture relationships between notes, while efficiently reducing sequence length.

\begin{figure*}[!htbp]
	\centering
    \includegraphics[width=0.8\linewidth]{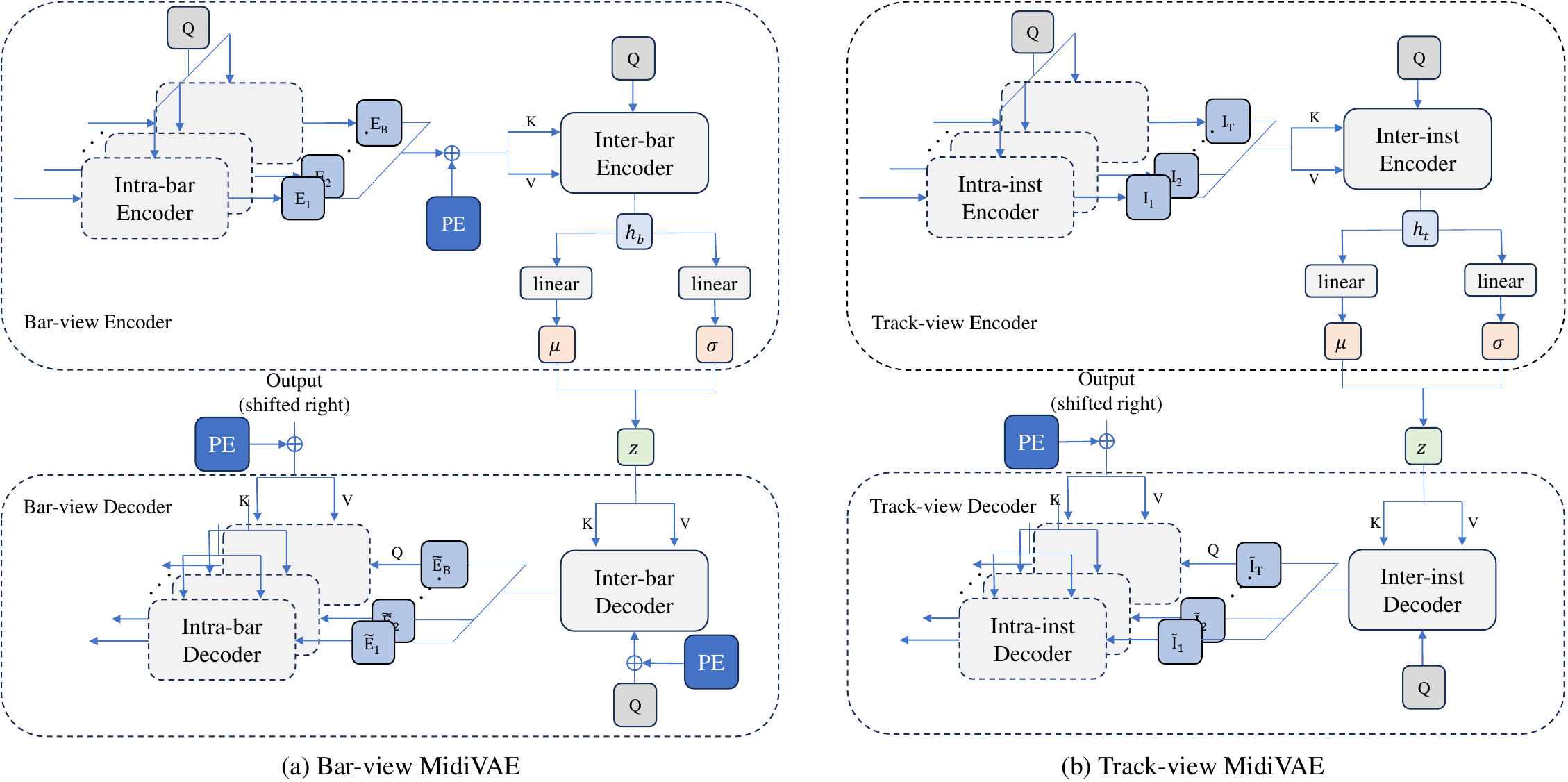}
    \vspace{-0.4cm}
	\caption{
	(a) The details of Bar-view MidiVAE, where ``PE" refers to the positional encoding.
	(b) The details of Track-view MidiVAE, 
 where ``Intra-inst" and ``Inter-inst" respectively denote Intra-instrument and Inter-instrument.
 }

	\label{fig:two-view}
	\vspace{-0.5cm}
\end{figure*}

\vspace{-0.2cm}
\subsection{Bar-view MidiVAE}
\vspace{-0.1cm}

Generating long music usually involves long feature sequences, which makes it difficult to model the details of the musical composition \cite{ji2023survey}.
For the purpose of allowing the model to fully learn the local details about the musical composition, we design Bar-view MidiVAE, which consists of a hierarchical structure of encoder and decoder. 

Fig.\ref{fig:two-view}(a) shows the details of the Bar-view MidiVAE. 
The Bar-view encoder is made up of two parts: a weight-sharing intra-bar encoder and an inter-bar encoder.
Both of them comprise of a transformer encoder and a multi-head attention module. 
The intra-bar encoder is designed to model the bar embeddings $\left \{ E_1,E_2,\dots,E_B | E_i \in \mathbb{R}^{D_b}, \ 1 \leq i \leq B \right \}$, where $D_b$ represents the dimension of the embedding and $B$ denotes the number of bars.
Due to the obvious temporal relationship between bars, we add positional encoding to the bar embeddings and then pass them into the inter-bar encoder to model the inter-bar relationships.
Finally, we utilize a learnable query parameter to capture bar-view embedding $h_b$.
There are two components inside the Bar-view decoder: a weight-sharing transformer decoder-based intra-bar decoder and a mutli-head attention-based inter-bar decoder. 
The inter-bar decoder regards $z$ as the key and value, and uses a learnable query parameter with positional encoding to model guidance bar embeddings
$\left \{ \widetilde{E}_1,\widetilde{E}_2,\dots,\widetilde{E}_B | \widetilde{E}_i \in \mathbb{R}^{D_b} \right \} $.
To guide the reconstruction of the corresponding subsequence, the guidance embeddings are fed to the intra-bar decoder one after another.
From the view of bar, Bar-view MidiVAE effectively model the finer details of notes within bars and the local information of the music composition. 

\vspace{-0.1cm}
\subsection{Track-view MidiVAE}
In multi-track music, there are multiple instruments played simultaneously.
Notes within each instrument collaborate with each other, while the instruments themselves perform independently yet harmoniously. 
Therefore, it is crucial to model the characteristics of each instrument and their relationships, which is the key to capturing global information of the musical composition \cite{ji2023survey}.

To this end, we design a Track-view MidiVAE that consists of a hierarchical structure of encoder and decoder, as shown in Fig.\ref{fig:two-view}(b).
The Track-view encoder consists of two components: a parameter-sharing intra-instrument encoder and an inter-instrument encoder, both of which consist of a transformer encoder and a multi-head attention module.
The intra-instrument encoder is utilized to model the instrumental characteristics $\left \{ I_1,I_2,\dots,I_T | I_j \in \mathbb{R}^{D_t} , \ 1 \leq j \leq T \right \}$, where $D_t$ represents the dimension of the characteristic and $T$ denotes the number of instruments.
These characteristics serve as the key and value for attention module, while there is a learnable query parameter to exploit associations between characteristics, and ultimately generates track-view embedding $h_t$.
Correspondingly, the decoder comprises two components: a parameter-sharing transformer decoder-based intra-instrument decoder and a multi-head attention-based inter-instrument decoder.
The inter-instrument decoder regards $z$ as the key and value, and uses a learnable query parameter to model guidance instrumental characteristics
$\left \{ \widetilde{I}_1,\widetilde{I}_2,\dots,\widetilde{I}_T | \widetilde{I}_j \in \mathbb{R}^{D_t} , \ 1 \leq j \leq T \right \} $.
Each guidance characteristic is then passed into the intra-instrument decoder to guide the reconstruction of the corresponding subsequence.
From the view of track, Track-view MidiVAE effectively model the instrumental characteristics and harmony as well as the global information of the musical composition.


\renewcommand{\arraystretch}{1.2}
\begin{table*}[!htbp]
\centering
\caption{The reconstruction performance measured in terms of accuracy: overall, duration, pitch, position, instrument, bar and tempo.}
\scalebox{0.9}{
\begin{tabular}{cccccccc}
\hline
\multirow{2}{*}{Model}& \multirow{2}{*}{$\mathrm{Overall}\uparrow$}& \multicolumn{6}{c}{Attributes}  \\ 
\cline{3-8}
& & $\mathrm{Duration}\uparrow$ & $\mathrm{Pitch}\uparrow$ & $\mathrm{Position}\uparrow$ &$ \mathrm{Instrument}\uparrow$ &$ \mathrm{Bar}\uparrow$ & $\mathrm{Tempo} \uparrow$\\
\hline
Multi-track MusicVAE (REMI+) & 0.8704 & 0.6660 & 0.6679 & 0.9832 & 0.9433 & 0.9993 & 0.9628\\
Multi-track MusicVAE (OctupleMIDI) & 0.9018 & 0.6818 & 0.7540 & 0.9845 & 0.9942 & 0.9993 & 0.9975\\
Track-view MidiVAE & 0.9482 & 0.8115 & 0.8968 & 0.9867 & 0.9956 & 0.9997 & 0.9991\\
Bar-view MidiVAE & 0.9450 & 0.8224 & 0.8740 & 0.9880 & 0.9863 & 0.9998 & 0.9994\\
Multi-view MidiVAE &  \textbf{0.9690} & \textbf{0.8940} & \textbf{0.9294} & \textbf{0.9909} & \textbf{0.9999} & \textbf{0.9999} & \textbf{0.9999}\\
\hline
\end{tabular}}
\label{tab:lab1}
\vspace{-0.5cm}

\end{table*}
\vspace{-0.2cm}
\subsection{Multi-view MidiVAE}
\vspace{-0.1cm}
The aforementioned Track- and Bar-view MidiVAE capture the global information,  instrumental characteristics and harmony as well as local details of the music composition, respectively.
For the purpose of leveraging the strengths of both models, we consider the musical composition from these dual views.

We first convert the original sequence $S$ into a Track-view sequence $S_t \in \mathbb{R}^{T\times L_t\times M}$ and a Bar-view sequence $S_b \in \mathbb{R}^{B\times L_b\times M}$,
where $L_t$ and $L_b$ represent the number of notes in a track and the number of notes in a bar, respectively, and $M$ means the number of musical attributes. 
The conversion process can be represented by the following equation:
\begin{equation}
    S_t = X_t \cdot S,  \ S_b = X_b \cdot S
\end{equation}
where $X_t$ represents the transformation matrix according to the track-based rule and $X_b$ represents the transformation matrix according to the bar-based rule.
Then $S_t$ and $S_b$ are passed into the Track- and Bar-view encoders to model $h_t$ and $h_b$, respectively. 

Given the multi-view information, we can implement a hybrid variational encoding-decoding strategy to map them to the same latent space and then reconstruct with an adaptive feature fusion method, as shown in Fig.\ref{fig:totalarch}. 
To be specific, $h_t$ and $h_b$ are concatenated together and fed into a Conv1d layer for multi-view information fusion to obtain the hybrid embedding, from which we can get $\mu$ and $\sigma$ of the latent distribution. 
We utilized reparameterization to obtain the latent representation $z \in \mathbb{R}^{D_z}$, which serves as the input of the Track- and Bar-view decoders to predict probability matrix $P_t$ and $P_b$ of two sequences.
To obtain the probability matrix $P$ of the reconstruction sequence, we utilize an adaptive weight vector $\alpha$ to fuse $P_t$ and $P_b$ by the following equation:
\begin{equation}
    P = \alpha \cdot X^{-1}_t \cdot  P_t + (1 - \alpha) \cdot
    X^{-1}_b \cdot P_b  
\end{equation}
During training, our model needs to reconstruct $S$, $S_t$ and $S_b$ simultaneously for loss computation.
We formulate our loss function as follows:
\begin{equation}
    \mathcal{L}_{\text{total}} = \mathcal{L}_{\text{rs}} + \mathcal{L}_{\text{rst}} + \mathcal{L}_{\text{rsb}} + \beta \cdot \mathcal{L}_{\text{KL}}
\end{equation}
$\mathcal{L}_{\text{rs}}, \mathcal{L}_{\text{rst}}$, and $\mathcal{L}_{\text{rsb}}$ represent the reconstruction loss of  $S$, $S_t$ and $S_b$, while $\mathcal{L}_{\text{KL}}$ stands for the Kullback-Leibler divergence and $\beta$ is the hyperparameter of $\beta$-VAE \cite{higgins2016beta}.

\vspace{-0.2cm}
\section{EXPERIMENTS}
\vspace{-0.2cm}
\label{sec:pagestyle}
\subsection{Dataset}
To train and evaluate our models, we utilize the CocoChorales dataset \cite{wu2022chamber}. 
The dataset consists of 240,000 pieces, and each of them is a standard four-part chorale (Soprano, Alto, Tenor, Bass) in $\frac{4}{4}$ time of 8 bars. 
The training set contains 192,000 pieces, whereas the validation and test sets each contains 24,000 pieces.
In total, there are 13 different instruments included in the dataset, namely Violin, Viola, Cello, Trumpet, French Horn, Trombone, Tuba, Flute, Oboe, Clarinet, Saxophone, Bassoon and Double Bass. 
Each sample consists of a combination of four of these instruments.

 \vspace{-0.1cm}
\subsection{Experiment Setup}
 \vspace{-0.1cm}
All models were trained using the Adam optimizer with an initial learning rate of 1e-4. 
We used a cross-entropy loss against the ground truth and teacher forcing for all models.

To testify the effectiveness of our improvements, the following models were compared.
Aiming at a fair comparison, the identical experimental setup was implemented for each model.
(1) \textbf{Multi-track MusicVAE}:
We chose Multi-track MusicVAE with REMI+ \cite{von2022figaro} as the baseline. 
For a fair evaluation, we considered 8 bars as an extended bar and fed it into the baseline.
Besides, a Multi-track MusicVAE using OctupleMIDI as the input was implemented. 
Following \cite{simon2018learning}, the hidden size of LSTM in encoder and decoder were $\left \{2048, 1024\right \}$. The dimension of $z$ is set to 512.
(2) \textbf{Track-view MidiVAE}: The layers of the transformer encoders and decoders were $\left\{4,8 \right\}$, while all multi-head attention modules employed 8 heads.
(3) \textbf{Bar-view MidiVAE}: For a fair comparison, the Bar-view MidiVAE used the same configurations as the Track-view MidiVAE in transformer blocks and multi-head attention modules.
(4) \textbf{Multi-view MidiVAE}: The configurations of the modules followed the Track- and Bar-view MidiVAE. Additionally, the hyperparameters were $D_b=D_t=D_z=512$.

We performed reconstruction and generation experiments to gauge the model performance. 
The reconstruction experiments are based on overall accuracy and attributes accuracy of sequence reconstruction, with the results presented in Table \ref{tab:lab1}.
For generating experiment, we organized a listening test involving 20 individuals instructed to listen to 20 audio samples generated by each model, and provided ratings for each sample based on three criteria: melody, instrumental harmony and overall quality. The mean opinion score (MOS) results are shown in Table \ref{tab:lab2}.

 \vspace{-0.2cm}
\subsection{Comparison with Baseline}
 \vspace{-0.1cm}
To compare our model with baseline, we performed reconstruction and generation experiments. 
Table \ref{tab:lab1} shows a significant improvement in the reconstruction accuracy of our purposed model, with the overall, duration and pitch accuracy improved by 9.9\%, 22.8\% and 26.1\%, respectively.
In addition, it also outperforms baseline in other attributes accuracy.
As can be seen from Table \ref{tab:lab2}, the music samples generated by Multi-view MidiVAE exceed the baseline's by 0.673 in overall MOS, 0.683 in melody MOS, and 0.655 in instrumental harmony MOS, achieving the best results in all metrics.
Both sets of experimental results demonstrate the effectiveness of our improvements in the usage of OctupleMIDI and Multi-view strategies.

 \vspace{-0.2cm}
\subsection{Ablation Study}
\subsubsection{The Effect of OctupleMIDI}
\vspace{-0.15cm}
With the purpose of demonstrating the effectiveness of the OctupleMIDI, we compare the Multi-track MusicVAE using inputs of REMI+ and OctupleMIDI.
Relative to REMI+, the usage of of OctupleMIDI results in a 75\% decrease in sequence length and a 357\% increase in the model's inference speed.
From results of Multi-track MusicVAE (REMI+) and Multi-track MusicVAE (OctupleMIDI) in Table \ref{tab:lab1} and Table \ref{tab:lab2} we can see that, the Multi-track MusicVAE using OctupleMIDI exhibits improved reconstruction and generation capabilities, illustrating the usage of OctupleMIDI can improve the model's ability to reconstruct and generate musical composition.

\begin{table}[h]
    \vspace{-0.2cm}
    \begin{center}
    \caption{The MOS on melody, instrumental harmony and overall quality of different models with 95\% confidence intervals. The “R” and
    “O” denote REMI+ and OctupleMIDI respectively.}
    \vspace{-0.2cm}
    \label{tab:lab2}
    \scalebox{0.81}{
        \begin{tabular}{cccc}
            \hline
            \multirow{2}{*}{Model}& \multirow{2}{*}{Melody} & \multirow{2}{*}{\shortstack{Instrumental\\Harmony }} & \multirow{2}{*}{Overall}\\
            & & & \\
            \hline
            Multi-track MusicVAE (R)  & 2.769$\pm$0.098 &  2.821 $\pm$ 0.097 & 2.785$\pm$ 0.092 \\
            Multi-track MusicVAE (O) & 2.948$\pm$ 0.090 & 2.993 $\pm$ 0.081  & 2.998$\pm$ 0.075\\
            Track-view MidiVAE & 3.224$\pm$  0.089& 3.398$\pm$ 0.091 &3.276$\pm$ 0.081\\
            Bar-view MidiVAE & 3.335$\pm$ 0.0954 &  3.229$\pm$ 0.101& 3.309$\pm$ 0.082\\
            Multi-view MidiVAE & \textbf{3.452}$\pm$ \textbf{0.089} &   \textbf{3.476}$\pm$ \textbf{0.086} & \textbf{3.458}$\pm$ \textbf{0.079} \\
            Ground truth & 3.671$\pm$ 0.926  &  3.726$\pm$ 0.089 & 3.74$\pm$ 0.082\\
            \hline
        \end{tabular}
        }
    \end{center}
    
    \vspace{-0.8cm}
\end{table}
 \vspace{-0.05cm}
\subsubsection{Investigation of Multi-view} \vspace{-0.15cm}
To demonstrate the effectiveness of Track- and Bar-view MidiVAE, we conduct further comparisons.
As shown in Table \ref{tab:lab1} and Table \ref{tab:lab2}, both of Track- and Bar-view MidiVAE show significant improvements in reconstruction accuracy and sample qualities relative to Multi-track MusicVAE. 
In particular, Track-view MidiVAE achieves higher results in pitch accuracy, overall accuracy and the instrumental harmony MOS of the generated samples, which shows that the Track-view MidiVAE exhibits a superior capability in modeling musical global information, instrumental characteristics and harmony.
While the Bar-view reflects a better performance of duration accuracy and position accuracy, illustrating that the Bar-view MidiVAE performs better in capturing local details.
In addition, we also compare Multi-view MidiVAE with Track- and Bar-view MidiVAE.
The results show that our proposed model yields significant improvements in both reconstruction accuracy and MOS of generated samples, demonstrating it effectively leverages the strengths of both dual views without conflict.

 \vspace{-0.25cm}
\section{CONCLUSIONS}
 \vspace{-0.15cm}
\label{sec:typestyle}
This paper introduces Multi-view MidiVAE to effectively model and generate long multi-track symbolic music.
The Multi-view MidiVAE utilizes a 2-D representation, OctupleMIDI, to capture relationships among notes while reducing the feature sequence length.
Furthermore, through a hybrid variational encoding-decoding strategy, we integrate both Track- and Bar-view MidiVAE features to concentrate on instrumental characteristics and harmony as well as global and local information about the musical composition.
Objective and subjective experimental results\footnote{Demo page: \href{https://zwlinzw.github.io/Multi-view_MIDIVAE_demo}{https://thuhcsi.github.io/icassp2024-Multi-view\_MIDIVAE}} on the CocoChorales dataset demonstrate that Multi-view MidiVAE is indeed a efficient VAE approach for modeling long multi-track symbolic music.

\textbf{Acknowledgements}: This work is supported by National Natural Science Foundation of China (62076144), Shenzhen Science and Technology Program (WDZC20220816140515001, JCYJ20220818\\101014030) and Shenzhen Key Laboratory of next generation interactive media innovative technology (ZDSYS20210623092001004).
\bibliographystyle{IEEEbib}
\bibliography{strings,refs}

\end{document}